\AtBeginDocument{%
	\providecommand\BibTeX{{%
			\normalfont B\kern-0.5em{\scshape i\kern-0.25em b}\kern-0.8em\TeX}}}

\newcommand{\name}{LLMTreeRec\xspace}

\newcommand{\ie}{\emph{i.e.,}\xspace}
\newcommand{\eg}{\emph{e.g.,}\xspace}
\newcommand{\etal}{\emph{et al.}\xspace}

\pdfoutput=1

\documentclass[11pt]{article}

\usepackage{coling}

\usepackage{times}
\usepackage{latexsym}

\usepackage[T1]{fontenc}

\usepackage[utf8]{inputenc}

\usepackage{microtype}

\usepackage{inconsolata}

\usepackage{graphicx}

\usepackage{booktabs} 
\usepackage{amsmath}
\usepackage{amsfonts}
\usepackage{mathrsfs}
\usepackage{amssymb}
\usepackage{subcaption}
\usepackage{multirow}
\usepackage[linesnumbered,ruled]{algorithm2e}
\let\oldnl\nl
\newcommand{\nonl}{\renewcommand{\nl}{\let\nl\oldnl}}
\usepackage{multicol}  
\usepackage{enumitem}
\setlist{nosep}
\usepackage{array}
\usepackage{float}
\usepackage{hyperref}
\usepackage{comment}
\usepackage{tcolorbox}  
\tcbuselibrary{breakable}   

\usepackage{color, xspace}
\usepackage{xcolor}
\usepackage{dialogue}

\definecolor{dialoguebg}{HTML}{F0F0F0}
\definecolor{usercolor}{RGB}{0,0,0}
\definecolor{llmcolor}{RGB}{0,128,0}

\newtcolorbox{dialoguebox}{
  colback=dialoguebg,
  sharp corners,
  boxrule=0pt,
  boxsep=0pt,
  left=10pt,
  right=10pt,
  top=5pt,
  bottom=5pt
}

\title{LLMTreeRec: Unleashing the Power of Large Language Models for Cold-Start Recommendations}

\author{
  \textbf{Wenlin Zhang\textsuperscript{1}},
  \textbf{Chuhan Wu\textsuperscript{2}},
  \textbf{Xiangyang Li\textsuperscript{2}},
  \textbf{Yuhao Wang\textsuperscript{1}},
  \textbf{Kuicai Dong\textsuperscript{3}}
  \\
  \textbf{Yichao Wang\textsuperscript{2}},
  \textbf{Xinyi Dai\textsuperscript{2}},
  \textbf{Xiangyu Zhao\textsuperscript{1}}$^*$,
  \textbf{Huifeng Guo\textsuperscript{2}},
  \textbf{Ruiming Tang\textsuperscript{2}}\thanks{Corresponding Authors}
\\
  \textsuperscript{1}City University of Hong Kong, Hong Kong, China \\
  \textsuperscript{2}Noah’s Ark Lab, China
  \textsuperscript{3}Noah’s Ark Lab, Singapore
\\
    \{wl.z, yhwang25-c\}@my.cityu.edu.hk, xianzhao@cityu.edu.hk
    \\
    \{wuchuhan1, lixiangyang34, dong.kuicai, wangyichao5
    \\
    daixinyi5, huifeng.guo, tangruiming\}@huawei.com
}

\begin{document}
\maketitle
\begin{abstract}
The lack of training data gives rise to the system cold-start problem in recommendation systems, making them struggle to provide effective recommendations.
To address this problem, Large Language Models(LLMs) can model recommendation tasks as language analysis tasks and provide zero-shot results based on their vast open-world knowledge. 
However, the large scale of the item corpus poses a challenge to LLMs, leading to substantial token consumption that makes it impractical to deploy in real-world recommendation systems. 
To tackle this challenge, we introduce a tree-based LLM recommendation framework \name, which structures all items into an \textit{item tree} to improve the efficiency of LLM's item retrieval.
\name achieves state-of-the-art performance under the system cold-start setting in two widely used datasets, which is even competitive with conventional deep recommendation systems that use substantial training data. Furthermore, \name outperforms the baseline model in the A/B test on Huawei industrial system. Consequently, \name demonstrates its effectiveness as an industry-friendly solution that has been successfully deployed online. Our code is available
at: \footnote{\label{foot1}\url{https://github.com/Applied-Machine-Learning-Lab/LLMTreeRec}}

\end{abstract}

\section{Introduction}
\label{sec:introduction}

Recommendation systems collect user behavioral data(\eg clicks, likes, pages viewed, and etc) to understand the preferences, historical choices, and characteristics of users and items~\cite{bobadilla2013recommender}, then provide personalized recommendation results.
Conventional recommendation systems require substantial user-item interaction data to capture collaborative information. However, real-world recommendation systems often face the cold-start challenge, where the lack of user-item interactions or insufficient training data leads to the recommendation system being unable to provide personalized recommendations. Specifically, the cold-start challenge can be categorized into (1) \textit{User cold-start}: Recommending for new users with limited history~\cite{huang2022learning, pandey2016resolving}, (2) \textit{Item cold-start}: Recommending new items with limited user interactions~\cite{pan2019warm, vartak2017meta}, (3) \textit{User-item cold-start}: Recommending for new users and new items~\cite{sanner2023large, li2019zero}, and (4) \textit{System cold-start}: Recommending under the assumption that no training set is available~\cite{hou2024large}.
Most of the existing works assume the training set is available, and mainly focus on either user cold-start or item cold-start problems. In this paper, we focus on the system cold-start problem(\ie provide recommendation results without any training set).

\begin{figure*}[t]
    \centering
    \includegraphics[width=0.85\linewidth]{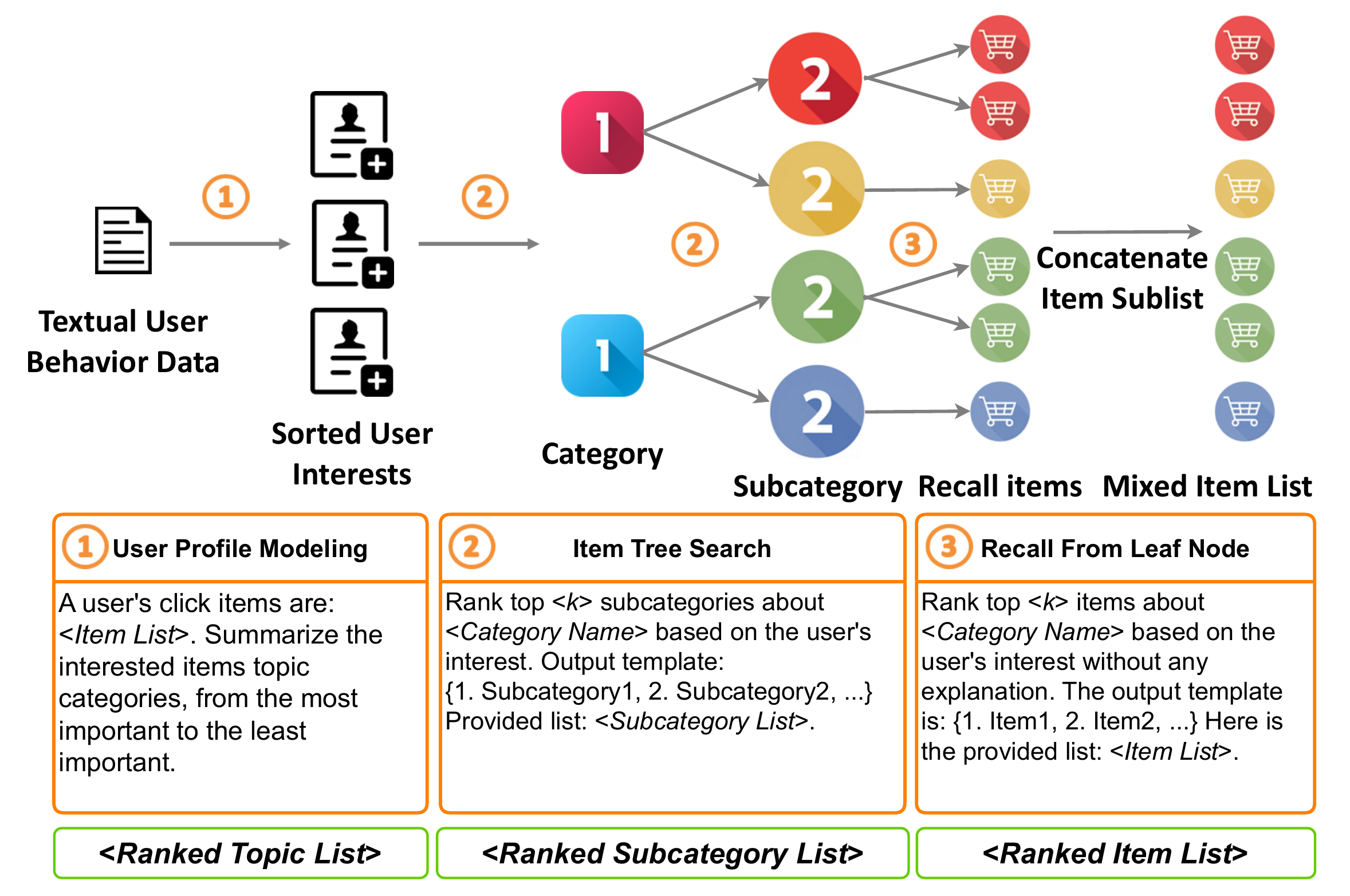}
    \caption{The overview of \name: LLM-centered tree-based recommendation framework.}
    \label{fig:Fig2_Framework}
    \vfill
\end{figure*}

\begin{figure*}[t]
\setlength\abovecaptionskip{0.2\baselineskip}
\setlength\belowcaptionskip{0.2\baselineskip}
    \centering
    \includegraphics[width=0.8\linewidth]{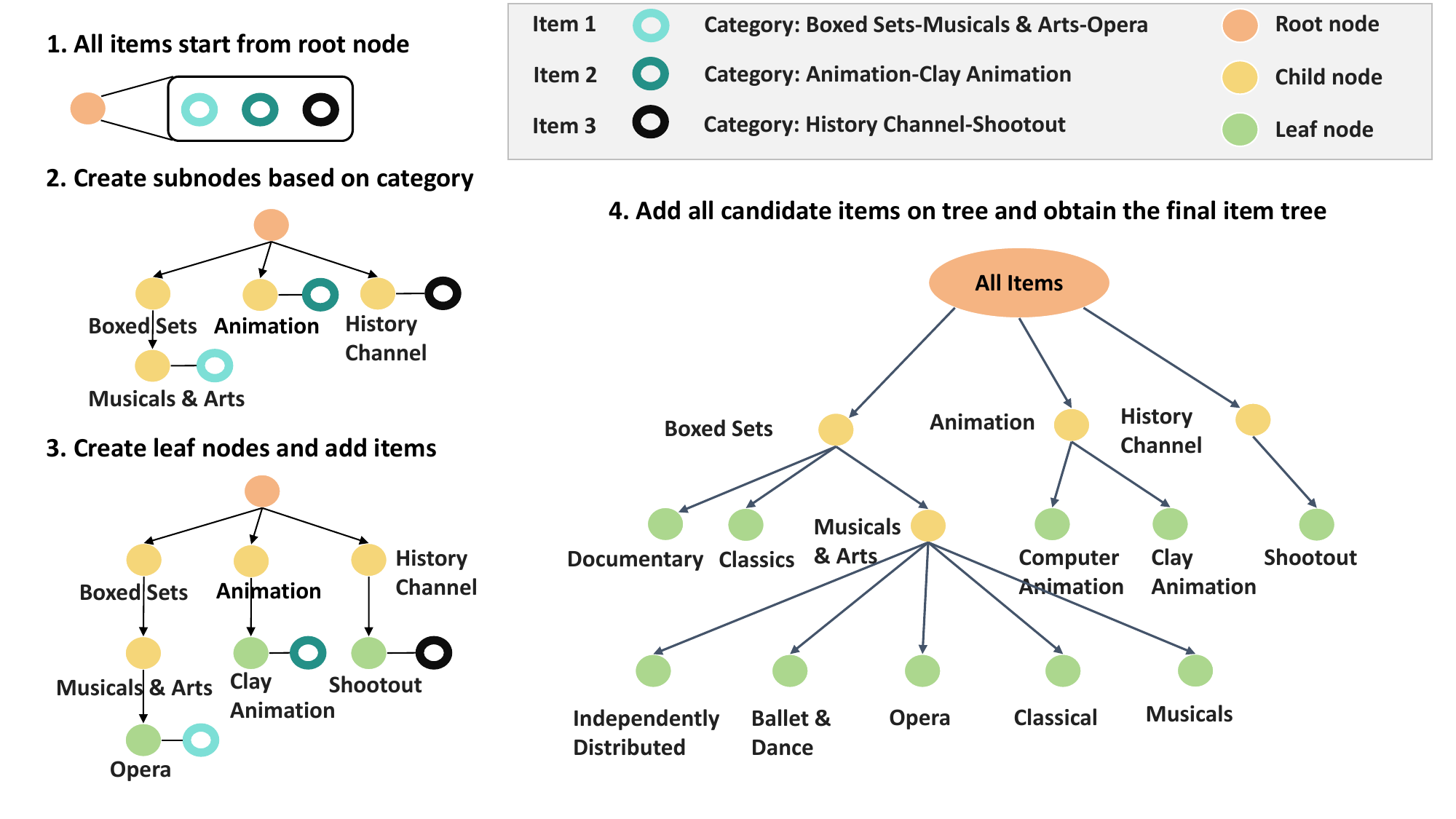}
    \caption{The procedure of item tree construction.}
    \label{fig:Fig3_ItemTree}
\end{figure*}

The recent emergence of Large Language Models (LLMs), such as ChatGPT~\cite{NEURIPS2020_1457c0d6} and Claude~\cite{bai2022constitutional}, has demonstrated robustness and generalization to excel in a broad spectrum of Natural Language Processing (NLP) tasks. The inherent potential of LLMs positions them as natural zero-shot solvers, which are capable of addressing cold-start recommendation challenges. 
In recent research by \citet{sanner2023large}, they collected a small dataset of item-based and language-based user preference data, based on which they validated that LLMs with only language-based preference show competitive performance with collaborative filter models under near cold-start settings. However, \textbf{challenge 1 arises:} LLMs lack the content understanding of candidate items. Based on the knowledge from pre-training corpora, LLMs may have a simple understanding of items, but there is still a gap between the general knowledge of LLMs and the domain-specific knowledge required in recommendation scenarios. To enable LLMs to incorporate the content information of items for recommendation without any training set available, it is necessary to input it into LLMs in text form. 
\citet{hou2024large} arranged items into a sequence form, using LLMs to rank small-scale items, and verified that LLMs can achieve competitive zero-shot ranking capabilities in system cold-start settings. However, \textbf{challenge 2 arises}: LLMs cannot simultaneously process all items in natural language form due to the input length limitation. 
As the scale of candidate items increases, the sequential input of item information will significantly increase the token requirement of the LLMs and interfere with the LLM's inference. Moreover, the massive item scale in recommendation system makes it infeasible to directly input all items into the LLM in natural language form.


To address the aforementioned challenges, we propose \name, a novel LLM-based framework that leverages large-scale item information for recommendations under the system cold-start setting. Specifically, we generate user preferences in natural language based on the user's interaction history and then leverage LLMs to recall items from a large-scale corpus.
To enable LLMs to handle large-scale item corpus, we have developed an innovative tree-based recall strategy. This involves constructing a tree that organizes items based on semantic attributes such as categories, subcategories, and keywords, creating a manageable hierarchy from an extensive list of items. Each leaf node in this tree encompasses a manageable subset of the complete item inventory, enabling efficient traversal from the root to the appropriate leaf nodes. Hence, we can recall items from the selected leaf nodes only. This approach sharply contrasts with traditional methods that require searching through the entire item list, resulting in a significant optimization of the recall process.

In summary, we highlight our contributions in three-fold:
\begin{itemize}[leftmargin=*]
    \item We propose \name, a novel LLM-centered tree-based recommendation framework that leverages user preferences and item information in natural language form, which can perform recommendations under the system cold-start setting. 
    \item We design a novel hierarchical item tree structure that can organize large-scale items into smaller, manageable segments contained within leaf nodes. The item tree can reduce the number of LLM input tokens required by 85\%, making it industry-friendly.
    \item \name achieves state-of-the-art performance under the system cold-start setting on two benchmark datasets, with its performance even competitive with conventional deep recommender models trained on substantial data. Furthermore, \name outperforms the baseline in A/B test on the Huawei system, and has been successfully deployed online.
\end{itemize}
\section{Proposed Framework}
\label{sec:framework}

In this section, we elaborate in detail on our proposed \name that can utilize large-scale item corpus information under the system cold-start setting. 
The overall framework of \name is depicted in Figure~\ref{fig:Fig2_Framework}.

\subsection{Item Tree Construction} \label{sec:sit}
Under the system cold-start setting, LLMs lack understanding of item information and require textual information inputs into the LLM. However, the large scale of the recommendation item corpus makes it difficult to input extensive item information into the LLM. 
To address these challenges, we propose the use of a hierarchical tree structure to organize items into leaf nodes. This approach facilitates LLMs to efficiently handle a large number of items. Figure~\ref{fig:Fig3_ItemTree} illustrates the construction procedure for an item tree.
Before outlining the construction of the item tree, we will introduce the formulated definitions for both the semantic information of an item $i$ and an item tree $\mathcal{T}$.
The semantic information of an item $i$ can be represented as $[s_1, s_2, \cdots, s_{k_i}]$, where $s_1, \cdots, s_{k_i}$ indicate semantic information in various levels (e.g., categories, subcategories, keywords, and other relevant details if needed), with granularity ranging from $s_1$ to $s_{k_i}$ in a coarse-to-fine manner. Here, $k_i$ represents the total number of semantic components of item $i$.
The item tree structure $\mathcal{T}$ organizes candidate items while each node in $\mathcal{T}$ corresponds to a subset of the item set. Specifically, the root node $root$ of the tree (depicted in red color) corresponds to a set encompassing all candidate items. Starting from the root node, items are partitioned into different child nodes $node_c$ (depicted in yellow color) based on their hierarchical semantic information. Each node on the tree corresponds to a set containing items with the same semantic prefix $[s_1, \cdots, s_j]$ where $j \leq k$.
The leaf nodes $node_l$ (depicted in green color) of the tree correspond to the smallest subsets into which each individual item $i$ can be categorized. After constructing the item tree, each specific item can be retrieved at its corresponding leaf node.

\subsection{LLM-Centered Tree-based Recommendation Framework} \label{sec:framework}
Based on item tree, we propose a novel LLM-centered tree-based recommendation framework (\name). 
To make LLMs comprehend user preferences and utilize item information of large-scale corpus, we design a chain-of-recommendation process, which enables the LLMs to retrieve information based on an item tree, referring to Section~\ref{ssec:model_rec}. Moreover, we elaborate on our effective retrieval strategy on item tree that enables \name to recall related items among large-scale item sets in Section~\ref{ssec:search}

\subsubsection{Chain-of-Recommendation Strategy} \label{ssec:model_rec}
With the aid of the item tree $\mathcal{T}$, we design a chain-of-recommendation strategy to integrate it with our recommendation process seamlessly.
\name provides an effective way for LLMs to handle large-scale item sets under the system cold-start setting. The recommendation chain in \name is executed in a single session, following the steps outlined below:


\paragraph{User Profile Modeling.}
Due to the privacy concerns about the user profile, we use the user's interaction history $H = [i_1, \cdots, i_{n_{u}}]$ for each user $u$ as LLM input for user profile modeling. Consequently, the user profile modeling function can be defined as:
\begin{equation}
    I = \text{UserProfileModeling}(H), 
\end{equation}
where $I$ is the inferred interest. 

            

User profile modeling and subsequent tasks are completed within the same session. As a result, LLMs are able to capture user preferences by leveraging the context history of user interactions stored in $H$ and the inferred interest $I$. This enables LLMs to successfully execute subsequent tasks in accordance with the user's preferences.



\paragraph{Item Tree Search.}
\name traverses the item tree from the root node to its child nodes. The search stops when the leaf node is reached. Each step deduces and ranks the top categories based on user interaction history and interest. More details are discussed in Section~\ref{ssec:search}. Formally, the item tree search function can be defined as
\begin{equation}
    childnodes = \\ \text{ItemTreeSearch}(H, I, node), 
\end{equation}
where $childnodes$ denotes the child node list selected by LLM.
LLM searches the child node list of $node$, infers based on the semantic information of child nodes, user interaction history $H$, and interests $I$ to give a ranked list of child nodes. 




LLMs iteratively perform item tree searches until reaching leaf nodes $node_l$.

\paragraph{Recall from Leaf Node.}
Every leaf node corresponds to a small subset of items that cannot be further divided based on semantic information. Hence, the text describing all items in the subset can be easily fed into \name. 
Then, \name will recall top items by considering user interaction history and interest.

Formally the function is defined as
\begin{equation}
    items = \text{RecallFromLeafNode}(H, I, subset, k), 
\end{equation}
where $items$ denotes the ranked recall items, item subset $subset$ is obtained from the corresponding leaf node $node_l$, and $k$ denotes the recall number from $subset$. The prompt template of Chain-of-Recommendation is illustrated in Figure~\ref{fig:Fig2_Framework}.

\begin{algorithm}[tb]
    \caption{\name}
    \label{alg:Alg1_UniLLMRec}
    \KwIn{User-item interaction history $H$}
    \KwOut{Recommended item list $L$}
    \nonl \textbf{Initialize}: $L = [], S = \text{Stack}()$

    Inferred interests: $I = \text{UserProfileModeling}(H)$ \\
    S.push(\textit{root})
    
    \While{$|L| < n$}{
        $node = S.\text{top}()$
        
        $S.\text{pop}()$
        
        \eIf{$\text{node is leaf node}$}{
            Get item $subset$ from $node$:
            $items = \text{RecallFromLeafNode}(H, I, subset, k)$
            
            $L.\text{add}(items)$
        
        }{
            $childnodes = \text{ItemTreeSearch}(H, I, node)$
            
            \For{$\text{node in childnodes.reverse()}$}{
                $S.\text{push}(node)$
            }
        }
    }

\end{algorithm}
\subsubsection{Search Strategy} \label{ssec:search}
The purpose of our search strategy is to balance between the diversity and relevance of retrieved items. To achieve fast retrieval to the target leaf node, we apply Depth-first Search (DFS) on our item tree. 
In particular, throughout each step of the search, only the top-ranked nodes will be selected for further DFS search, allowing \name to bypass less relevant nodes.
Upon reaching a leaf node, \name will recall the top $k$ items from the item subset of this leaf node.
The search ends if either (i) all leaf nodes are traversed, or (ii) the desired number of $n$ items has been recalled.
The parameter $k$ effectively serves a lever to modulate the diversity of recalled items.
Opting for a smaller $k$ increases the recommendation diversity, but at the cost of increased search time. Conversely, a larger $k$ tends to reduce diversity while expediting the search process.
The detailed pipeline of UniLLMRec is demonstrated in Algorithm~\ref{alg:Alg1_UniLLMRec}. 

\section{Experiment}
\label{sec:experiment}
We investigate three research questions(RQ): (1) How does the performance of zero-shot \name competitive to traditional models trained on different fractions of the training set? (2) How much does \name reduce the token requirements for LLM handling large-scale items? (3) How do the hyper-parameter and prompt design impact the recommendation result of \name? 

We will first introduce the experiment setting in Section~\ref{sec:experiment_setting}, then display the performance comparison experiment in Section~\ref{sec:sec_performance}, demonstrate the token consumption analysis in Section~\ref {sec:token_analysis}, show experiments about hyper-parameter and prompt design in Section~\ref{sec:hyper_parameter} and Section~\ref{sec:prompt_study}, and finally provide case study in Section~\ref{sec:case_study}.



\subsection{Experiment Setting}\label{sec:experiment_setting}

\begin{table}[t]
    \begin{tabular}{cccc}
    \toprule
    Dataset & Training set & Test set & Candidates \\
    \midrule
    MIND    & 51,283            & 500 & 1217   \\
    Amazon  & 70,728            & 500 & 6176   \\
    \bottomrule
    \end{tabular}
    \caption{The statistic detail of dataset.}
    \label{tab:tab1_datasets}
\end{table}

\subsubsection{Datasets}
In the experiments, we utilized two benchmark datasets including the MIND dataset~\cite{wu2020mind} and Amazon Review dataset~\cite{he2016ups} in the category of Movies and TV. 
Since handling the extensive item subsets from some leaf nodes posed challenges for direct input into the language model using a single prompt template, we constrained each subset in leaf node to a maximum of 50 items. Positive items were grouped into their respective subsets, and negative sampling was applied to each leaf node until reaching a size of 50. This process resulted in a candidate set of 1217 items for MIND and 6176 items for Amazon.

To ensure fair comparisons in our experiments, the length of user-item interaction sequences is truncated to 50, and all methods exclusively utilized the item titles as features. We randomly select 500 samples as testing set for both two dataset and list the statistics in Table~\ref{tab:tab1_datasets}.

\subsubsection{Evaluation Metrics}
We focus on evaluating the performance of the proposed framework and baseline in recall and re-ranking tasks. For each model, we primarily consider its Recall metric and the Normalized Discounted Cumulative Gain (NDCG) in the top-$k$ recall task.

\subsubsection{Baselines}
\name are compared with Popularity-based recommendation, FM~\cite{rendle2010factorization}, DeepFM~\cite{guo2017deepfm}, NRMS~\cite{wu2019neural}, SASRec~\cite{kang2018self}, and LLM-Ranker~\cite{hou2024large}.

\subsubsection{Implementation Details}

\name leverages \textbf{gpt-3.5-turbo}\footnote{\label{foot2}\url{https://platform.openai.com/docs/models/gpt-3-5}} and \textbf{gpt-4-1106-preview}\footnote{\label{foot3}\url{https://platform.openai.com/docs/models/gpt-4-1106-preview}} as the backbone LLM. Due to budget constraints, experiments on the \name method with GPT-4 as the backbone focus on the comparison of recall and NDCG metrics, which is elaborated in Section~\ref{sec:sec_performance}.

The constructed item tree structure shows difference in MIND and Amazon.
The item tree depth in MIND dataset is 2, with all leaf nodes merely located in the second layer. There are 17 and 276 nodes in the first and second layers respectively. 
As for the Amazon dataset, items without titles or semantic information are discarded during item tree construction. Subsequently, the constructed tree has a depth of 4, and the leaf nodes may be located in all layers. The node numbers from the first layer to the fourth layer are 78, 298, 126, and 19, respectively. 
In the item tree search stage, we set the recall subnode number as 10. Meanwhile, in the experiments, the parameter $k$ in the recall stage serves to limit the number of selected leaf nodes and is set to 5.

For popularity-based models, we select the most popular 20 items as a recommendation list. 
For LLM-Ranker, gpt-3.5-turbo is used as the backbone model.
Since LLM-Ranker cannot direct handle large-scale items, we randomly sample 100 items as the candidate item set for LLM-Ranker.
For conventional models, the FM, DeepFM, NRMS, and SASRec models all adopt a two-tower structure, utilizing the Adam optimizer with learning rate of 0.001.
FM uses TF-IDF (Term Frequency-Inverse Document Frequency)~\cite{salton1988term} of item titles as features, while in DeepFM, NRMS, and SASRec, item word embeddings are employed as features. We use a negative sampling ratio of 1 across all models. In the MIND dataset, all models only use the item title as the input feature. In the Amazon dataset, they use the item description as the input feature.
Finally, we increase the 10\% training set size for each model until the model performance is equivalent to \name. Thus, we can evaluate the performance between the capabilities under system cold-start setting and supervised 
conventional recommendation models.

\begin{table*}[]
\centering
\begin{tabular}{cccccc}
\toprule
\multicolumn{2}{c}{\multirow{2}{*}{Model}}        & \multicolumn{2}{c}{MIND}                                          & \multicolumn{2}{c}{Amazon}                                        \\
\multicolumn{2}{c}{}                              & Recall@20                       & NDCG@20                         & Recall@20                       & NDCG@20                         \\ 
\midrule
\multirow{4}{*}{Trained}   & FM                   & 0.0125                          & 0.0016                          & 0.0020                          & 0.0013                          \\
                           & DeepFM               & 0.0367                          & 0.0037                          & 0.0060                          & 0.0009                          \\
                           & NRMS                 & \textbf{0.0525}                 & 0.0306                          & 0.0660                          & 0.0105                          \\
                           & SASRec               & 0.0124                 & 0.0059                          & \underline{0.0800} & \underline{0.0162} \\
\midrule
\multirow{4}{*}{Zero-shot} & Pop                  & 0.0012                          & 0.0004                          & 0.0002                          & 0.0001                          \\
                           & LLM-Ranker           & 0.0213                          & 0.0201                          & 0.0040                          & 0.0040                          \\
                           & LLMTreeRec (GPT-3.5) & 0.0296                          & \underline{0.0359} & 0.0120                          & 0.0047                          \\
                           & LLMTreeRec (GPT-4)   & \underline{0.0509} & \textbf{0.0619}                 & \textbf{0.0964}                 & \textbf{0.0741}                 \\ 
    \bottomrule
\end{tabular}
\caption{Performance Comparison on two benchmark datasets. The conventional recommender systems including FM, DeepFM, NRMS, and SASRec are trained by 20\% training set.}
\label{tab:tab2_performance}
\end{table*}

\subsection{Performance Comparison (RQ1)} \label{sec:sec_performance}
The overall performance of \name and baselines are shown in Table~\ref{tab:tab2_performance}.
Specifically, the proposed \name framework is compared with the methods in two categories:

\begin{itemize}[leftmargin=*]
    \item The first is the method under the system cold-start setting.
    The popularity-based method, hampered by the absence of user-specific information, demonstrated an exceedingly low recall of items. LLM-Ranker outperforms popularity-based methods in both Recall and NDCG metrics, yet it lags behind \name(GPT-3.5) and \name(GPT-4). \name is capable of selecting the candidate set based on user interests and item trees, resulting in a refined candidate set compared to LLM-Ranker, thereby leading to improved performance.
    \item The others are the conventional recommendation models with training sets. Our main focus lies in evaluating how the performance of \name is competitive with conventional recommendation models with varying amounts of training data. Table~\ref{tab:tab2_performance} reports the results of conventional recommender systems with 20\% training set.
\end{itemize}

In summary, both \name(GPT-3.5) and \name(GPT-4), which do not require training, achieve competitive performance compared with conventional recommendation models that require training. Additionally, the \name(GPT-4) method significantly outperforms the \name(GPT-3.5) approach.

\subsection{Token Requirement Analysis (RQ2)}\label{sec:token_analysis}

\name recalls items from subsets based on the item tree, which effectively reduces the model's token requirement in the recall stage. We conduct a statistical analysis on the size of the candidate item set and the average token length for each item. After sampling, the MIND dataset comprises 1,217 items, while the Amazon dataset has 6,167 items, with an average token length of 14 and 10, respectively. The total tokens needed to input all items into the LLM exceeds ten thousand. \name effectively reduces the token requirement by item tree-based search. Figure~\ref{fig:Fig13_TokensConsumption} illustrates average token consumption in each framework stage.
Regarding input consumption, Stage 3 (\textit{Recall From Leaf Node}) of \name inputs the item IDs of all the retrieved leaf nodes into the LLM, resulting in high token requirements, accounting for over 50\% of the total. As for output consumption, since Stage 3 only recalls a limited number of items, all three stages have relatively low token consumption. In summary, the item tree-based search consumes minimal tokens, making \name a cost-efficient retrieval method.

\subsection{Hyper-parameter Analysis (RQ3)}\label{sec:hyper_parameter}

\begin{figure}[t]
\setlength\abovecaptionskip{0.2\baselineskip}
\setlength\belowcaptionskip{0.2\baselineskip}
  \centering
  \includegraphics[width=0.8\linewidth, trim=0 15 0 0,clip]{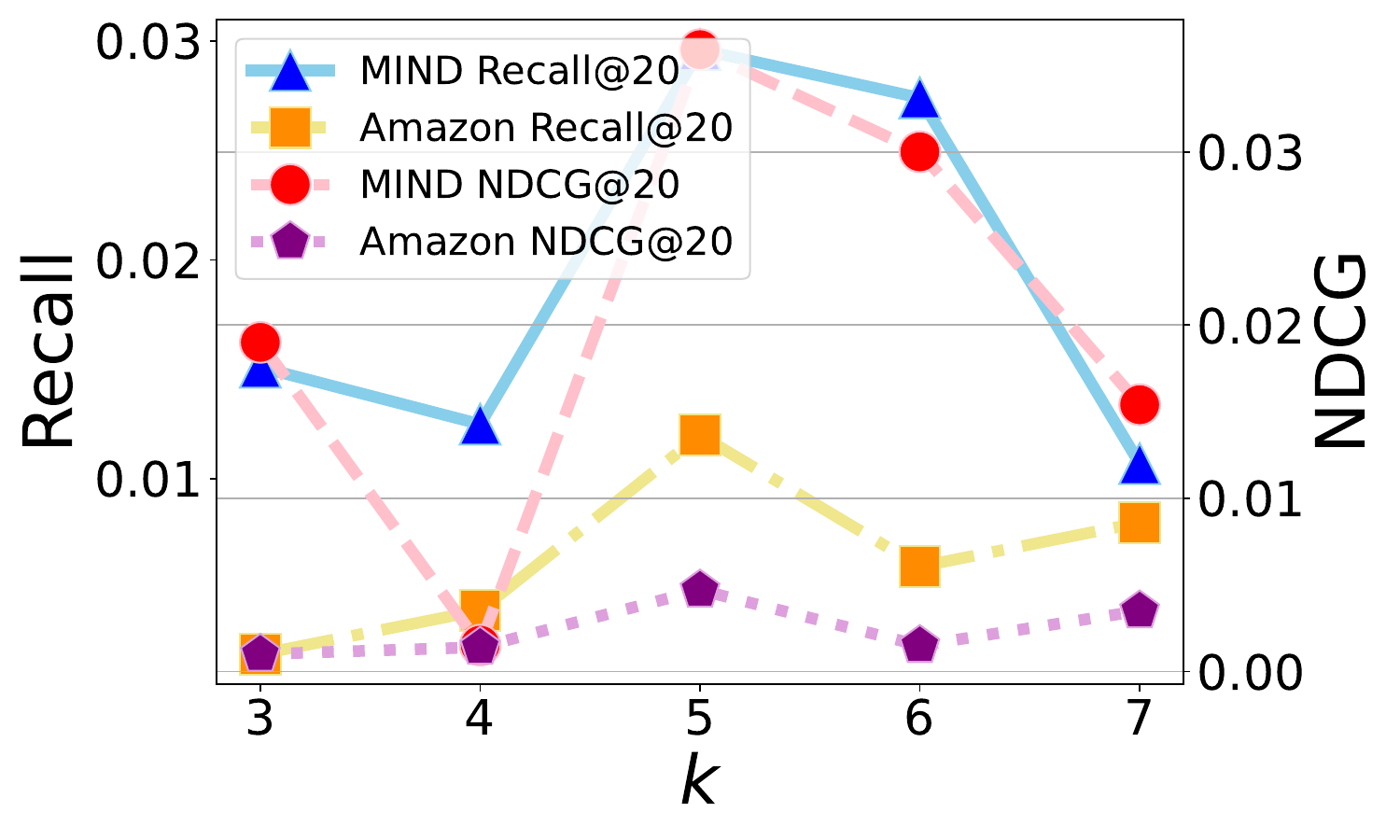}
  \caption{The impact of $k$ on recall performance.}
  \label{fig:Fig9_Parameter}
\end{figure}

The recall number $k$ in leaf nodes is the only hyper-parameter in \name. We conducted a study on the impact of $k$ on the recall task, and illustrate the results in Figure~\ref{fig:Fig9_Parameter}. As the value of $k$ increases, the number of items recalled by our model from different leaf nodes steadily rises. We observe a phenomenon where both recall rate and NDCG initially rise and then decline with the increasing $k$. Clearly, with the continuous increment of $k$, the number of items recalled from each node also increases, indicating that the model tends to recommend items from subsets that are of higher interest to the user. When $k$ decreases, the model recalls items from more leaf nodes, resulting in higher diversity in the retrieved results. In summary, the parameter $k$ plays a crucial role in the model by influencing the trade-off between diversity and the quantity of recalled items under different categories.  

\begin{figure}[t]
\setlength\abovecaptionskip{0.2\baselineskip}
\setlength\belowcaptionskip{0.2\baselineskip}
  \centering
  \includegraphics[width=0.8\linewidth, trim=0 10 0 0,clip]{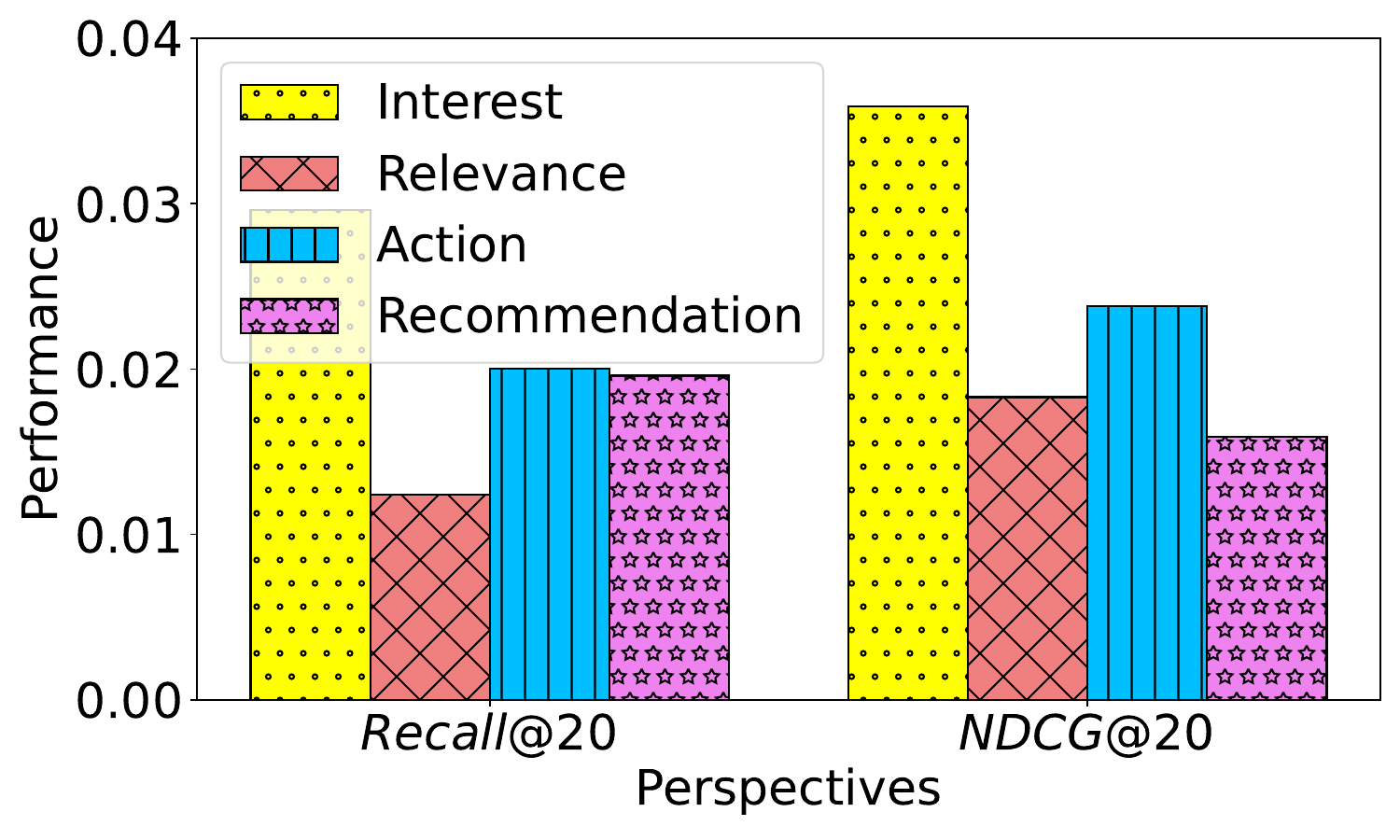}
  \caption{The impact of various perspective prompt design on recall performance.}
  \label{fig:Fig10_PromptDesign}
\end{figure}



\subsection{Prompt Study (RQ3)}\label{sec:prompt_study}

\begin{table*}[]
\centering
\small
\setlength\abovecaptionskip{0.2\baselineskip}
\setlength\belowcaptionskip{0.2\baselineskip}
\begin{tabular}{p{1.2cm{\centering}} p{3.6cm{\centering}} p{3.6cm{\centering}} p{3.6cm{\centering}}}
\toprule
Perspective     & User Profile Modeling & Item Tree Search  & Recall from Leaf Node \\
\midrule
Prompt template & A user's click items are: \textless{}Item List\textgreater{}. \textcolor{blue}{\textless{}Perspective-Variable   Prompt\textgreater{}}, from the most important to the least important. & Rank the top \textless{}k\textgreater subcategories about \textless{}Category Name\textgreater based on   \textcolor{blue}{\textless{}Perspective-Variable Prompt\textgreater} from the following candidates without any   explanation. The output template is: \{1. Subcategory1, 2. Subcategory2, ...\}   Here is the provided list: \textless{}Subcategory List\textgreater{}. & Rank the top \textless{}k\textgreater items about \textless{}Semantic Information\textgreater based on \textcolor{blue}{\textless{}Perspective-Variable Prompt\textgreater} from the candidates about \textless{}Topic\textgreater without any explanation. The output template is: \{1. Item1, 2. Item2, ...\}   Here is the provided list: \textless{}Item list\textgreater{}. \\
interest        & Summarize the interested items topic categories                                                                                                                       & the user's interest                                                                                                                                                                                                                                                                                                                                             & the user's interest \\
relevance       & Summarize the news topic categories related to users                                         & the relevance related to the user                                                                                              & the relevance related to the user                                                                         \\
action          & Summarize the news topic that the user are likely to click on                                                                                                         & the probability that the user is likely to click                                                                                                                                                                                                                                                                                                                & the probability that the user is likely to click                                                          \\
recommen-
dation  & Summarize the news topic worth recommending to the user                                    & the degree of recommendation to the user                                                                                       & the degree of recommendation to the user \\
\bottomrule
\end{tabular}
\caption{Prompt design from 4 various perspectives.}
\label{tab:tab2_prompt_study}
\end{table*}

We craft prompt templates from four different perspectives including interest, relevance, action, and recommendation tailored to the news recommendation as in Prompt4NR~\cite{zhang2023prompt}. The prompts designed from various perspectives are detailed in Table~\ref{tab:tab2_prompt_study} where the blue variant prompts are changed based on perspective. The performance of models under these four types of prompt settings is shown in Figure~\ref{fig:Fig10_PromptDesign} from which we can see that prompt design significantly impacts the model performance. Using a relevance-based prompt yielded a recall rate of only 1.24\% and an NDCG of 0.0183. By contrast, models using prompts of action and recommendation achieve approximately 2\% recall rate. Besides, the best performance is observed under the interest-based prompt design, where the recall rate and NDCG were twice that of the relevance prompt model.

These results underscore the significance of prompt design on non-fine-tuned LLMs in recommendation tasks. The interest-based prompt design can effectively leverage the LLM's ability to uncover user interests, thereby enhancing the personalization and precision of recommendations. 



\begin{figure}[t]
  \centering
  \includegraphics[width=0.9\linewidth, trim=0 10 0 0,clip]{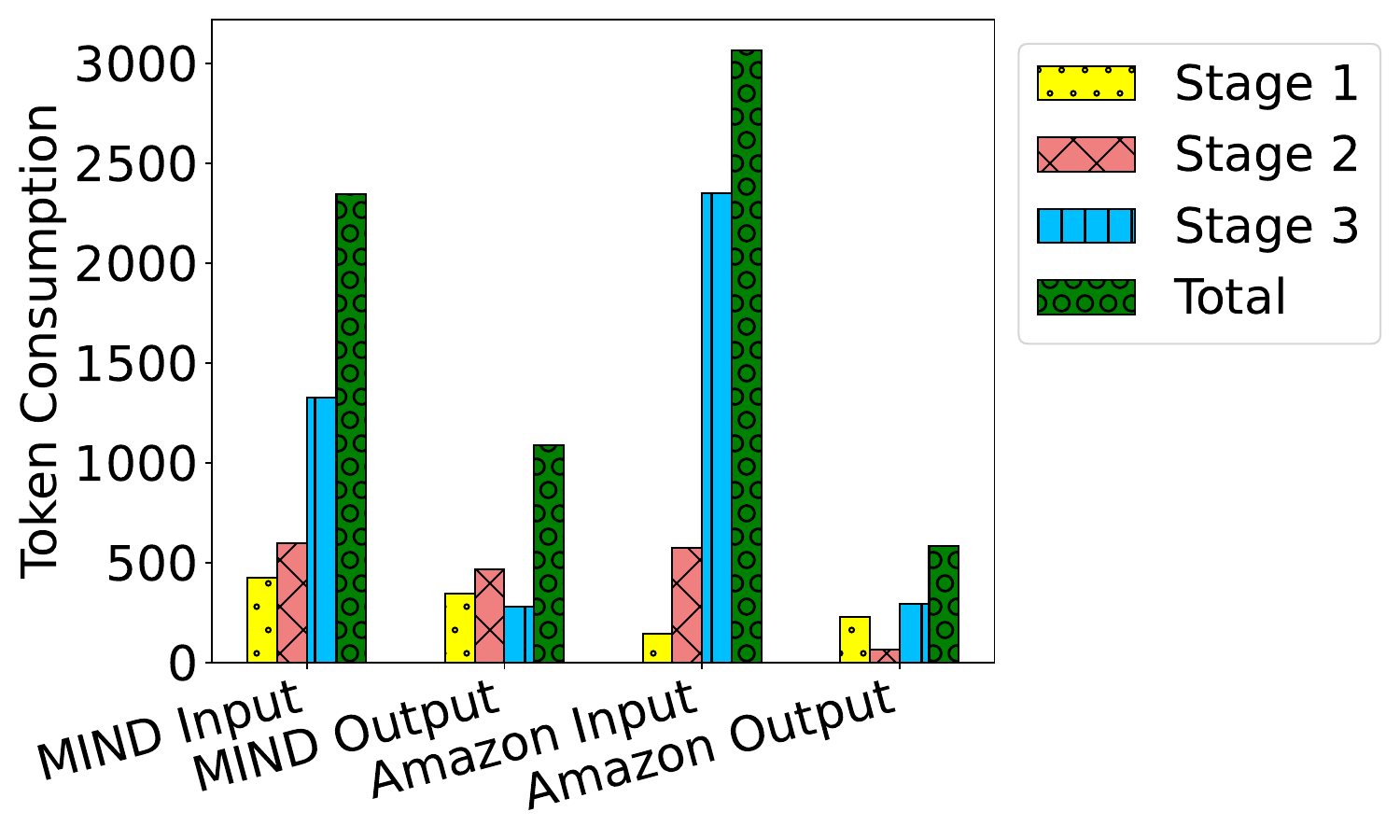}
  \caption{Consumption of tokens for each stage.}
  \label{fig:Fig13_TokensConsumption}
\end{figure}

\subsection{Case Study}\label{sec:case_study}
\name(GPT-3.5) or \name(GPT-4) can smoothly complete the entire pipeline procedure in most cases. However, both have a few bad cases in the experiment: (1) Even if the output format is clarified in the prompt template, but LLM does not strictly follow the instructions to output the item, resulting in the item not being indexed correctly.
(2) The challenge of hallucinations. In the user profile modeling stage, LLM summarizes user interests and provide representative examples of interest-related items. However, there is a hallucination risk where these examples might be erroneously included during the recall from leaf node.


\section{Industrial Online Performance}
\begin{table}[t]
\centering
    \begin{tabular}{cccc}
    \toprule
    Baseline & \name & Improvement \\
    \midrule
    0.577    & 0.725 & 25.64\%      \\
    \bottomrule
    \end{tabular}
    \caption{Performance comparison in Huawei recommendation system.}
    \label{tab:tab2_online_performance}
\end{table}

We compare \name with the
baseline model in online Huawei recommender systems to perform seven day A/B test. This scenario belongs to infomercial recommendations, where 1188 unique items are recalled from the library and eventually displayed in the user's terminal. The compared results on 662 sessions are reported in Table~\ref{tab:tab2_online_performance}. \name(backbone Huawei LLM) has achieved a substantial improvement in \textit{NDCG@10}, reaching 0.725 compared to the baseline's 0.577, resulting in a significant gain of 25.64\%. The online A/B test results have validated the superior performance of \name. It provides an efficient solution for handling large-scale items in LLM under the system cold-start setting.

\section{Related Work}
\label{sec:related_work}

\subsection{Cold-Start Recommendation}
The cold-start problem is a common challenge in recommender systems. The existing research mostly focused on addressing the user cold-start~\cite{huang2022learning, pandey2016resolving} and item cold-start problems~\cite{pan2019warm, vartak2017meta}: the models learn from the user-item interaction history and perform recommendations for new users or new items. To tackle this problem, many existing works have enhanced the embedding quality of users and items by incorporating side information~\cite{wang2019multi, yin2017spatial} or using pre-training models~\cite{li2019zero, hao2021pre}.
Different from the user/item cold-start problem, the system cold-start problem arises when recommender systems have no prior recommendation knowledge about users and items. LLMs can leverage their general knowledge and have achieved promising zero-shot performance on various natural language tasks, suggesting their potential to address the system cold-start problem~\cite{zhang2021language, hou2024large, ding2021zero}. However, LLMs struggle to handle large-scale item corpora due to the high computational cost involved during the ranking task.

\subsection{Large Language Model for Recommendation}
In recent years, large language models (LLMs) have shown their great potential and strong capability in handling different tasks like information retrieval~\cite{jia-etal-2024-mill, xu2024large}, medical prediction~\cite{10.1145/3626772.3657722, 10.1145/3627673.3679673}, knowledge graph completion~\cite{xu-etal-2024-multi}, knowledge distilation~\cite{MAAD}, and
computer vision~\cite{yang2023dawn}. 
Although the existing works~\cite{liang2023mmmlp, li2022gromov, liu2023multi, liu2023linrec, liu2024multimodal, wang2023plate, zhang2024glint} have made significant improvements to recommendation systems, the integration of LLMs with recommendation systems~\cite{10.1145/3627673.3679743, liullm, zhang2024notellm, liu2024large} can greatly enhance their performance.
In the recommendation community, existing methods~\cite{survey} incorporating LLMs can be categorized into two groups. On the one hand, some works directly generate the recommendation result of item ID \cite{geng2022recommendation,cui2022m6, hua2023index}. For example, P5 \cite{geng2022recommendation} reformulates recommendation tasks to natural language processing tasks utilizing personalized prompts and conducts conditional text generation. Hua \etal examine various item IDs based on P5~\cite{hua2023index}. Although LLMs' strong language understanding ability can promote the exploitation of text features in recommendation, LLMs that merely utilize the item embedding underexploit the collaborative information.
Item ID alignment with conventional recommender systems has received widespread attention recently. For example, CTRL and FLIP~\cite{ctrl,flip} encode the information of user-item pair by two embedding towers including a semantical model that encodes the textual feature and a collaborative model that processes the same sample in tabular form. Subsequently, the embeddings from two embedding towers are aligned via contrastive learning. 

To enable LLMs to acquire recommendation knowledge in system cold-start settings, some approaches~\cite{unictr, liu2023chatgpt, hou2024large} adopt a straightforward strategy of inputting the candidate set directly into the LLM. Moreover, in-context learning can provide several samples as the recommendation knowledge reference~\cite{liu2023chatgpt}. However, these methods face the challenge of high computational cost when dealing with large-scale item information.

\section{Conclusion}
\label{sec:conclusion}

We propose \name, an LLM-centered tree-based recommendation framework to tackle the system cold-start challenge. 
To deal with large-scale item sets, we design a novel strategy to structure all items into a hierarchical tree structure, \ie \textit{item tree}.
Based on the item tree, LLM effectively refines the candidate item set by utilizing this hierarchical structure for search.
Extensive experiments on MIND and Amazon datasets indicate that \name can achieve competitive performance compared to conventional recommendation models under the system cold-start setting.
Furthermore, \name is industry-friendly and easy to deploy on industrial recommender systems.

\section*{Acknowledgements}
This research was partially supported by Huawei (Huawei Innovation Research Program), Research Impact Fund (No.R1015-23), APRC - CityU New Research Initiatives (No.9610565, Start-up Grant for New Faculty of CityU), CityU - HKIDS Early Career Research Grant (No.9360163), Hong Kong ITC Innovation and Technology Fund Midstream Research Programme for Universities Project (No.ITS/034/22MS), Hong Kong Environmental and Conservation Fund (No. 88/2022), and SIRG - CityU Strategic Interdisciplinary Research Grant (No.7020046).
\bibliography{9Reference}




\end{document}